\documentstyle[aps,prb,multicol,epsf]{revtex}
\begin{document}

\title{The $\bf T=0$ $\bf 2k_F$ density wave phase transition
in a two dimensional Fermi liquid is first order}
\bigskip
\author{B. L. Altshuler}
\address{NEC Research Institute, 4 Independence Way, Princeton, NJ 08540 \\
Physics Department, MIT, Cambridge, MA 02139}
\author{L. B. Ioffe}
\address{Physics Department, Rutgers University, Piscataway, NJ 08855 \\
and Landau Institute for Theoretical Physics, Moscow}
\author{A. J. Millis}
\address{AT\&T Bell Laboratories, Murray Hill, NJ 07974}

\maketitle

\begin{abstract}

We study $T=0$ spin density wave transitions in two dimensional Fermi liquids
in which  the ordering wavevector $\bf Q$ is such that the tangents to the
Fermi line at the points connected by $\bf Q$ are parallel (e.g. $Q=2p_F$ in a
system with a circular Fermi line) and the Fermi line is not flat.
We show that the transition is first order if the ordering wave vector
$\bf Q$ is not commensurate with a reciprocal lattice vector, $\bf G$, i.e.
${\bf Q} \neq {\bf G}/2$.
If $\bf Q$ is close to ${\bf G}/2$ the transition is weakly first order and
an intermediate scaling regime exists; in this regime the $2p_F$ susceptibility
and observables such as the NMR rates $T_1$ and $T_2$ have scaling forms which
we determine.

\end{abstract}

\begin{multicols}{2}

\section{Introduction}
Quantum phase transitions have attracted substantial recent interest.
Antiferromagnet-singlet transitions in insulating magnets
\cite{Chakravarty,Ioffe,Sachdev,Chubukov}
and ferromagnetic and antiferromagnetic transitions in Fermi liquids
\cite{Hertz,Millis} have been studied in detail, and the crossover
between the insulating and fermi liquid critical points in two
spatial dimensions has also been studied \cite{Sachdev95}.
Here we consider an important case which has not so far been discussed in the
literature, namely what we call the ``$2p_F$'' spin or charge density wave
transition of a fermion system.
By ``$2p_F$'' we mean an ordering wavevector $\bf Q$ which connects two points
on the Fermi  line with parallel tangents [See Fig. 1] \cite{remark}.
For a circular Fermi line any vector $\bf Q$ of magnitude $2p_F$ connects
two such points.
In this paper we consider explicitly the spin density wave case, but our
results can be applied with only minor modifications to the charge density
wave case.
We assume that the Fermi line is not straight.
We also assume that Fermi liquid theory adequately represents the non-critical
properties of the fermions.
If it does not, our results do not apply.
We briefly discuss one non-Fermi liquid scenario in the conclusion.
One important motivation for studying the $2p_F$ case is the high $T_c$
superconducting material $La_{2-x} Sr_x CuO_4$, in which strong magnetic
fluctuations have been observed \cite{Aeppli}; the fluctuations are peaked at
an $x$-dependent wavevector $Q(x)$ which is claimed to be a ``$2p_F$''
wavevector of the Fermi line calculated by standard band-structure
techniques for this material \cite{Littlewood}.
Our results may also be relevant for quasi two dimensional materials such as
TTF-TCNQ \cite{Clark94}.

In order to study critical phenomena analytically, one expands about a mean
field solution.
If Fermi liquid theory is a good starting point, then
the appropriate mean field theory is the Random Phase
Approximation (RPA), in which the
susceptibility $\chi(\omega,{\bf q})$ is given in terms of the interaction
constant $g$ and the polarizability of noninteracting fermions,
$\Pi_0(\omega, {\bf q})$ by
$\chi(\omega,{\bf q})=\Pi_0(\omega,{\bf q})/(1-g^2\Pi_0(\omega,{\bf q}))$.
The transition occurs  at the wavevector $\bf Q$ as $g$ is increased
to the point that
$g^2 \Pi_0(0,{\bf Q})=1$.
In three spatial dimensions, $\Pi_0(0,{\bf q})$ is not maximal at any $2p_F$
wavevector, moreover $d\Pi_0(0,{\bf q})/dq$ is logarithmically divergent as
$q \rightarrow 2p_F$.
Therefore a ``$2p_F$'' transition is impossible in $d=3$ and so we focus on
$d=2$ in this paper.

In two dimensional case, $\Pi_0(0,{\bf q})$ is so strongly peaked at $q=2p_F$
that it is natural to assume that the spin density instability happens at
``$2p_F$''.
The ``$2p_F$'' case  is difficult to treat by the methods used previously to
study phase transitions at other momenta \cite{Hertz,Millis}.
In these works the fermions are ``integrated out'' and the problem is reduced
to a model of interacting bosonic spin fluctuations.
In the ``$2p_F$'' case the action functional obtained by integrating out
the fermions  has coefficients which are singular and non-analytic because
the  fermion response functions are non-analytic for $Q=2p_F$.
These non-analyticities lead to divergences in the action as $T \rightarrow 0$
and make it difficult to apply the conventional approach \cite{Hertz,Millis}.
Instead, in this paper we apply a perturbative renormalization group technique
to a model which includes both spin fluctuations and fermions.

Our perturbation parameter is $1/N$, the fermion spin degeneracy.
The leading order of the perturbation theory is the familiar RPA approximation.
The next order is a theory of electrons interacting by exchanging RPA
fluctuations.
We show that this theory is infrared divergent.
We sum the leading infrared divergent contributions using the renormalization
group.

The behavior of spin fluctuations changes dramatically if their wave vectors
are close to half of a reciprocal lattice vector, $\bf G$.
The important parameter is $\Delta G = |{\bf Q}-{\bf G}/2|$.
If $\Delta G$ is sufficiently small we must distinguish two regimes in
the renormalization group flow: large momenta, where the infrared cutoff is
greater than $\Delta G$ and small momenta, where it is less.
For large momenta the divergences are logarithmic; the logarithms may be
summed by renormalization group to power laws and we  use the
$1/N$-expansion to find the exponents.
Although  the physical value of $N=2$, the small value of the numerical
coefficients in front of these logarithms suggests that the exponents
obtained in the first order in $1/N$ are close to their exact values at
$N=2$.
For small momenta, the divergences are much stronger and, we show, drive the
transition first order as soon as the regime of small momenta is reached.

\begin{figure}
\centerline{\epsfxsize=6cm \epsfbox{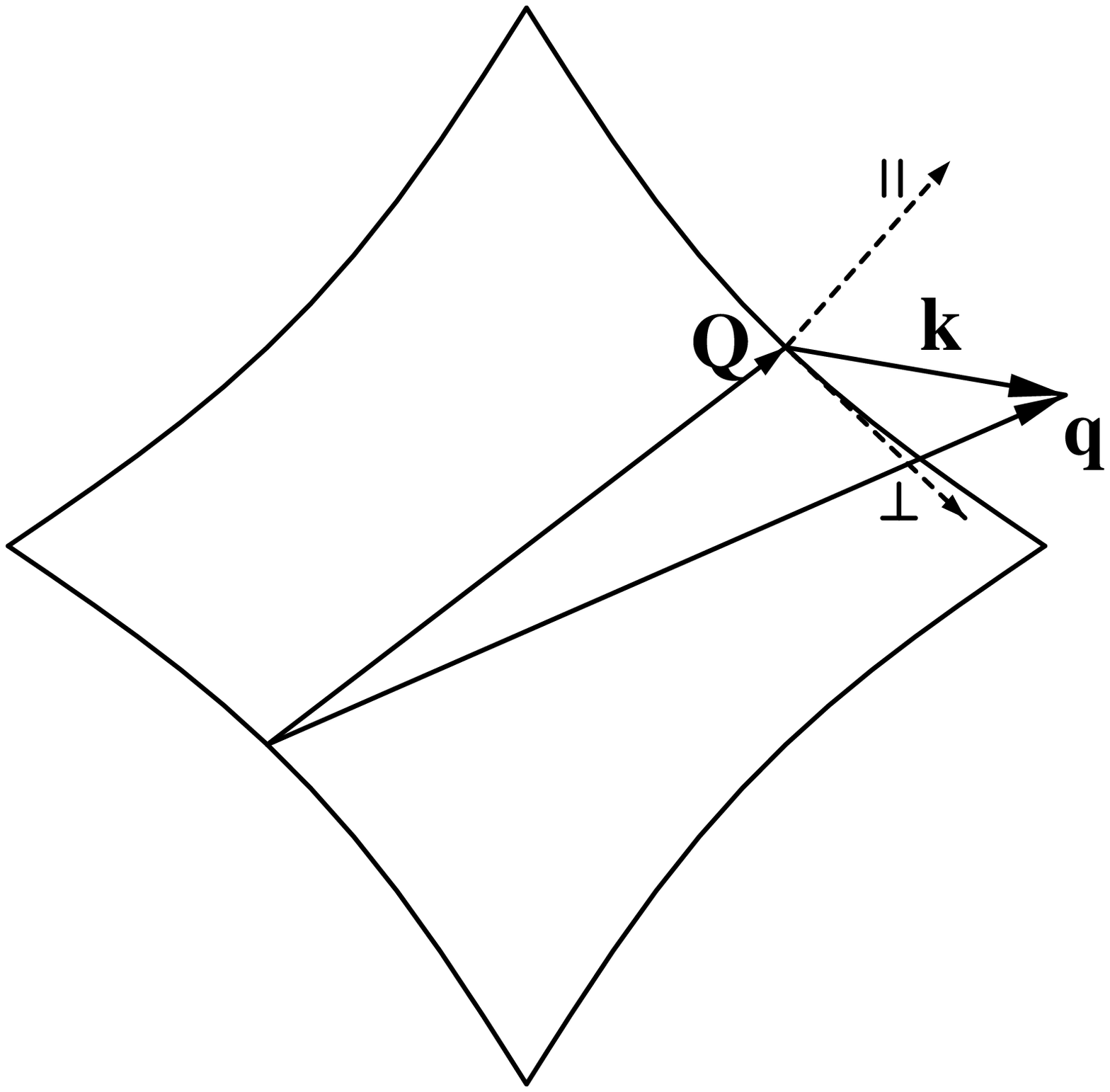}}
{
FIG 1.
Sketch of Fermi line and important wavevectors.
The Fermi line shown here is similar to that claimed  to be appropriate to
$La_{1.8} Sr_{.14} Cu O_4$.
The ordering wavevector $\bf Q$ for spin fluctuations connects two points on
the Fermi line.
It is assumed that the tangent to the Fermi line at one end of the vector
$\bf Q$  is parallel to the tangent to the Fermi line at the other end.
We have also shown a typical momentum of a spin fluctuation $\bf q$.
We parametrize the vector ${\bf k}={\bf q}-{\bf Q}$ by its cartesian
components $k_\perp$ and $k_\parallel$ in the coordinate system (shown by
dashed lines) associated with the Fermi line at the points connected by
${\bf Q}$.
}
\end{figure}
\vspace{0.2in}

The outline of this paper is as follows.
In Section II we define the model, derive the RPA theory and make a convenient
scaling of  variables.
In Section III we analyze the fluctuation corrections and derive
renormalization group equations in the regime of large momenta.
In Section IV we derive analogous equations in the regime of small momenta and
show that they imply that the transition is first order.
In Section V we discuss the physical consequences of our results.
Section VI contains a summary of the results, a discussion of their relation to
previous work on quantum critical phenomena and correlated electrons, a note
on the extension of our results to a charge density wave transition and a
conclusion.

\section{Model and Random Phase Approximation}

Our starting point is a Hamiltonian, $H$, describing fermions moving in a
lattice and interacting with each other via a short range four fermion
interaction $W$:
\begin{equation}
H = \sum_{p \alpha} \epsilon(p) c^\dagger_{p,\alpha} c_{p,\alpha} +
	W \sum_{p,p',q,\alpha,\beta} c^\dagger_{p,\alpha} c_{p+q,\alpha}
		c^\dagger_{p',\beta} c_{p'-q,\beta}
\label{H}
\end{equation}

We assume that a $T=0$ spin density wave transition to a state with long range
order at wave vector $\bf Q$ occurs as $W$ is increased to a critical value
$W_c$.
Because we expect the physics in this region to be determined by the exchange
of spin density fluctuations we use a Hubbard-Stratonovich transformation to
recast Eq. (\ref{H}) as a theory of fermions coupled to spin fluctuations $\bf
S_{q}$.
The theory is  described by the action
\begin{eqnarray}
{\cal A}\{c, S \} & = & \sum_{p \alpha} G^{-1}(\epsilon,p)
c^\dagger_{\epsilon,p,\alpha} c_{\epsilon,p,\alpha}
\nonumber \\
	&+& \sum_{\omega,q} D^{-1}(\omega,q)  {\bf S}_{\omega,q}
		{\bf S}_{-\omega,q}
\label{A} \\
	&+&
	g_b \sum_{p,q,\epsilon,\omega} c^\dagger_{\epsilon,p,\alpha}
	\vec{\sigma}_{\alpha \beta} c_{\epsilon+\omega,p+q,\beta}
	\vec{S}_{-\omega,-q}
\nonumber
\end{eqnarray}
Here $G(\epsilon,p)$ is the fermion Green function, $D(\omega,q)$ is the spin
fluctuation propagator and $g_b$ is a bare coupling constant derived from $W$.
When Hubbard-Stratanovich transformation is applied to Eq. (\ref{H}), the
result is $D_b(\omega,q)=1$, $g_b^2=W$ and $G_b(\epsilon,p)$ is the
non-interacting fermion Green function, i.e.
\begin{equation}
G_b( \epsilon , p) =
\frac{1}{i\epsilon - \epsilon(p) }
\label{G_bare}
\end{equation}
The interaction between spin fluctuations and fermions changes the form of the
fermion Green function and spin fluctuation propagator.
We assume that the effects of the short scale fluctuations which do not become
singular at the critical point can be described by  conventional Fermi liquid
renormalizations.

The leading order of perturbation theory for the action (\ref{A}) is the
Random Phase Approximation (RPA) which takes into account the renormalization
of the spin propagator by the electron polarization bubble,
$D_0^{-1}(\omega,q)=D_b^{-1}(\omega,q)-\Pi_0(\omega,q)$.

We shall be interested in momenta close to the momentum $Q$ at which
$\Pi_0(0,q)$ is maximal.
For wavevectors near $Q$ the momentum and frequency dependences of
$\Pi_0(0,q)$ are non-analytic and controlled by Fermi line
singularities.
Because the singular behavior of $\Pi_0(\omega,q)$ is controlled by the
distance from $\bf q$ to the Fermi line,
it will be convenient to parametrize the momentum $\bf q$ in terms of the
variables $k_\parallel$ and $k_\perp$ shown in Fig. 1.

The fermion polarizability $\Pi_0(\omega,q)$ can be calculated by
summing all diagrams which are irreducible with respect to the fermion-fermion
interaction and have two  external $S_{\omega, {\bf q}}$ legs.
This generalizes the RPA by including Fermi liquid corrections.
This sum has contributions from short length scale processes which give
$\Pi_0(\omega,q)$ an analytic dependence on $q$ and $\omega$ and also
contributions from Fermi line singularities, which lead to a
non-analytic dependence of $\Pi_0(\omega,q)$ on $q$ and $\omega$.
Thus we write $\Pi_0(\omega,q)=\Pi^{anal}(\omega,q)+\Pi^{sing}(\omega,q)$,
within Fermi liquid theory singularities come from the diagram shown in Fig. 2.
The analytic expression corresponding to this diagram is
\begin{equation}
\Pi^{sing}(\omega,q) = - g_0^2 \sum_{\epsilon,p}
G(\epsilon+\omega,p+q)G(\epsilon,p)
\end{equation}
where $g_0$ represents the interaction constant renormalized by Fermi liquid
corrections.
To obtain the form of Fermi line singularities we expand the spectrum of the
fermions in the vicinity of the points $\pm {\bf Q}/ 2$,
obtaining for the fermion Green function
\begin{equation}
G( \epsilon, p) =
\frac{1}{i z \epsilon - v_F p_{\parallel} + \frac{v_Fp_{\perp}^2}{2p_0}}
\label{G_expanded}
\end{equation}
Here $v_F$ is the renormalized Fermi velocity, $p_0$ is the radius of
curvature of the Fermi line and $p_\parallel (p_{\perp} )$ are momentum
components normal (tangential) to the Fermi line as measured from the points
$\pm {\bf Q}/2$ and $z$ is the quasiparticle residue.
Note that coordinates $p_\parallel$ and $p_{\perp}$ are compatible with the
spin fluctuation coordinates $k_\parallel$, $k_\perp$.

\begin{figure}
\centerline{\epsfxsize=4cm \epsfbox{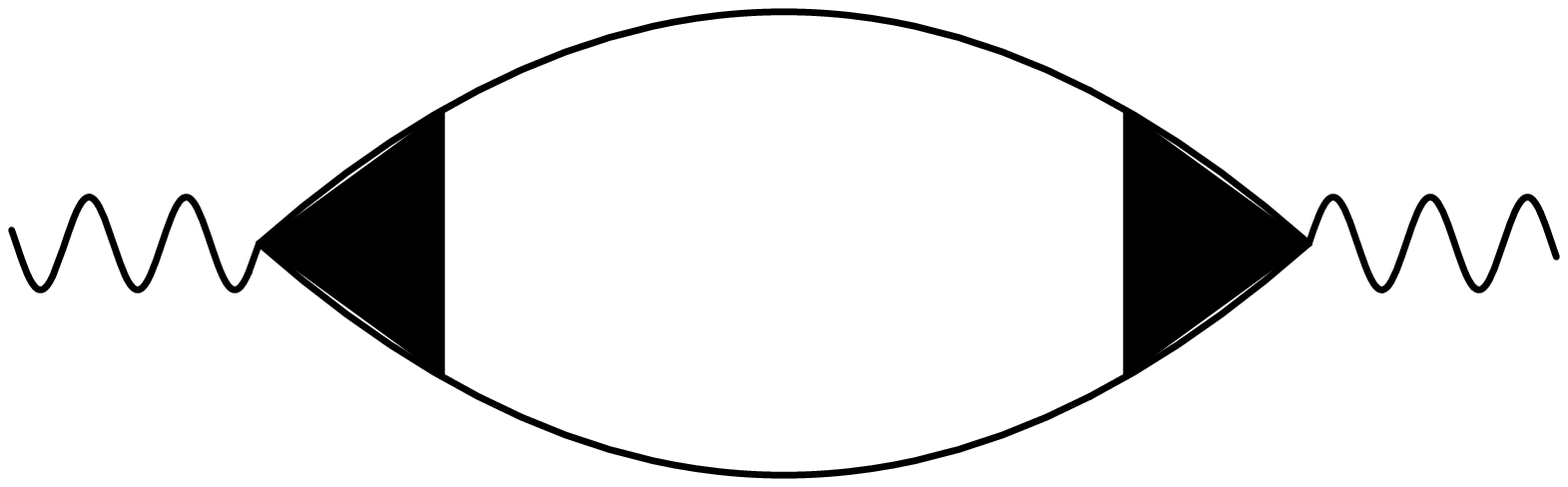}}
{
FIG. 2.
Diagram yielding the non-analytic momentum and frequency dependence
of the polarizability $\Pi_0(\omega,q)$ in Fermi liquid theory.
The solid lines are fermion propagators.
}
\end{figure}
\vspace{0.2in}

All dominant infrared contributions come from processes in which an
electron is scattered from one small region of fermion momenta around $\bf
Q/2$  to another around $-{\bf Q}/2$.
It is convenient to introduce dimensionless momenta parameterizing these
regions
and rescale all fields so that the resulting action does not contain
dimensionful variables.
We choose
\begin{equation}
\begin{array}{rclrcl}
p_\perp &\rightarrow& \sqrt{p_0 p_F} p_\perp & \hspace{0.5in}
c_{\epsilon,p} &\rightarrow& v_F^{-1} p_F^{-5/2} p_0^{-1/2} c_{\epsilon,p} \\
p_\parallel &\rightarrow& p_F p_\parallel &
S_{\omega,k} &\rightarrow& g_0^{-1} p_0^{-1/2} p_F^{-3/2} S_{\omega,k} \\
\epsilon &\rightarrow& v_F p_F \epsilon &
g_0 &\rightarrow& 1
\end{array}
\end{equation}
After this transformation we may assume we are dealing with a circular Fermi
surface of unit radius but with an upper cut off on angular integrals of
the order of $p_F/p_0$ and an upper cut off on radial integrals of the order
of $1$; the dimensionless parameter controlling the incommensurability becomes
\[
\Delta G \rightarrow \frac{|{\bf G}|}{2p_F} - 1
\]

In the new variables the action retains the general form (\ref{A}) but the
Green function of the fermions  changes to
\begin{equation}
G( \epsilon , p) =
\frac{1}{i z \epsilon - p_{\parallel} + p_{\perp}^2/2}
\label{G}
\end{equation}
while the bare spin fluctuation propagator becomes
\[
D_b(\omega,q)=g^2 \frac{\sqrt{p_0p_F}}{v_F}
\]

Evaluating the diagram shown in Fig. 2 yields the singular part of the
fermion polarizability in $d=2$
\begin{eqnarray}
\Pi^{(sing)}(\omega, {\bf q}) = - \frac{N}{2 \pi z}
	 Re \sqrt{k_\parallel+\frac{1}{4}k_\perp^2 - i \omega}
\label{Pi^(sing)}
\end{eqnarray}
Here and below we are use Matsubara frequencies $\omega=2\pi n T$ so that
$\Pi(\omega,{\bf q})$ is purely real.
$\Pi^{(sing)}(\omega, {\bf q})$ depends only on the combination
$k_\parallel+\frac{1}{4}k_\perp^2$ because for a circular Fermi line
$\Pi^{(sing)}(\omega, {\bf q})$ is a function only of $|\bf q|-2p_F$.

For electrons on a lattice there are additional images of
$\Pi^{(sing)}(\omega, {\bf q})$ coming from fermion transitions with momenta
shifted by reciprocal lattice vectors.
The most important of these is the transition
with momentum $\bf G-q$ which might be also close to $\bf Q$ in a typical
situation.
Generally, we expect that the singular contribution to fermion polarizability
comes from the transitions with momenta transfer $\bf q$ and $\bf G-q$:
\begin{eqnarray}
\Pi_0(\omega, {\bf q}) &=& \Pi^{(sing)}(\omega, {\bf q}) +
	\Pi^{(sing)}(\omega, {\bf G-q})
\nonumber \\
	&+& \Pi^{(anal)}(\omega, {\bf q})
\label{Pi_0}
\end{eqnarray}
Here $\Pi^{(anal)}(\omega, {\bf q})$ is analytical contribution to fermion
polarizability coming from fermion momenta far from ${\bf Q/2}$, its
dependence on the momenta and frequency is negligible relative to the strong
dependence coming from the singular parts.

There are two regimes of $\bf q$ at $T=0$.
For $k \gg \Delta G$, $\Pi_0(\omega, {\bf q})$ has a
symmetric square root peak at a wavevector indistinguishable from ${\bf G}/2$.
At smaller scales, $k \ll \Delta G$, the peaks separate and
each peak acquires asymmetric form: $\Pi(0,k) \sim -\sqrt{k}$ at $k>0$ and
$\Pi(0,k) \sim -|k|/(\Delta G)^{1/2}$ if $k<0$.
The qualitative form of $\Pi(0,k)$ is shown in Fig. 3.

\begin{figure}
\centerline{\epsfxsize=8cm \epsfbox{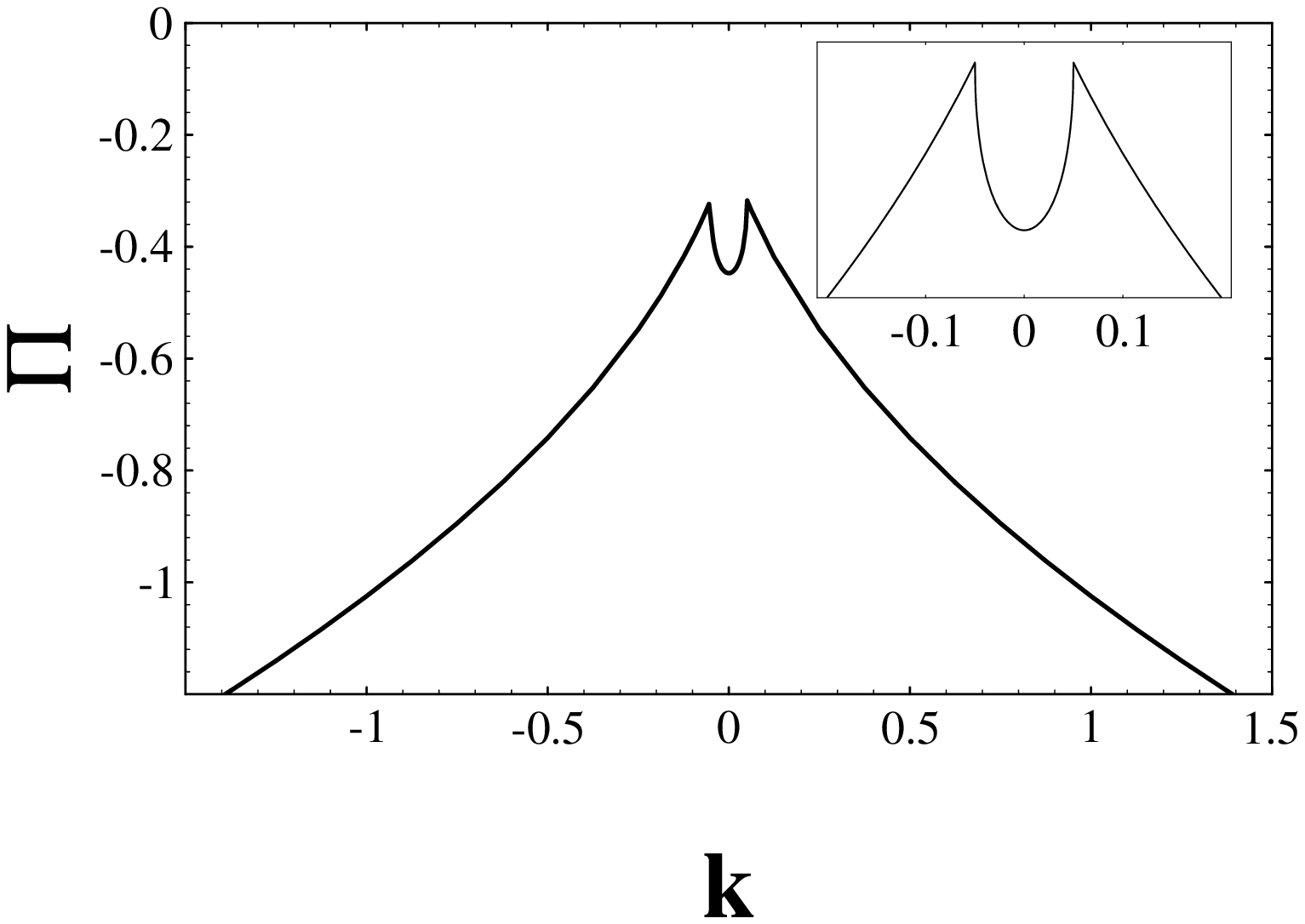}}
{
FIG. 3.
The qualitative form of fermion polarization $\Pi(0,k)$ in RPA approximation.
$k=0$ corresponds to the commensurate vector ${\bf G}/2$, peaks occur at
$k$ corresponding to ${\bf q}={\bf Q}$ and ${\bf q}={\bf G}-{\bf q}$.
The inset shows the blown up structure of the double peak maximum.
}
\end{figure}
\vspace{0.2in}

After RPA corrections are included, the spin fluctuation propagator entering
the action (\ref{A}) becomes
\begin{equation}
D_0(\omega,{\bf q})=\frac{1}{D_b^{-1}-\Pi_0(\omega, {\bf q})}
\label{D_0}
\end{equation}
Within the RPA, a second order phase transition occurs when the interaction
constant $g_0$ controlling the value of $D_0^{-1}$ is increased through the
critical value
\[
g_c^2= \frac{v_F}{\sqrt{p_0p_F} \Pi(0,0)}
\]

For $g=g_c$, $D(\omega,{\bf q})$ diverges as ${\bf q} \rightarrow {\bf Q}$
and $\omega \rightarrow 0$.
For $g \leq g_c$ the divergence is cut off.
It is convenient to define a bare dimensionless cutoff, $\Delta_0$ by
\begin{equation}
D_0(0,{\bf Q})=\frac{1}{\Delta_0}
\label{Delta}
\end{equation}
Within the RPA, $\Delta_0 \propto g_c-g$.

Corrections to the RPA involve diagrams in which electrons interact by
exchange of spin fluctuations.
These corrections may be organized in a $1/N$ expansion because the spin
fluctuation propagator is proportional to $1/N$ and each fermion loop contains
a factor of $N$.
The leading diagrams in $1/N$ are the self energy correction shown in Fig. 4
and the vertex correction shown in Fig. 5.
These diagrams are infrared divergent; the divergence is logarithmic if the
external momentum is larger than $\Delta G$ and power law if the external
momentum is less than $\Delta G$.
These two cases require separate discussions.


\section{Scaling at large momenta}

At large momentum the small difference between the ${\bf Q}$ and ${\bf G}/2$
is unimportant and we may write
\begin{eqnarray}
&D& \! (\omega,{\bf k}) =
\label{D}
\\
& & \frac{1}{\frac{g^2 N}{2\pi z}
	Re \left[ \sqrt{\frac{k_\perp^2}{4}+iz\omega+k_\parallel} +
	\sqrt{\frac{k_\perp^2}{4} + iz\omega - k_\parallel} \right] + \Delta}
\nonumber
\end{eqnarray}
for the spin fluctuation propagator.
We use this and the fermion Green function (\ref{G}) to calculate leading
corrections to the fermion self energy, the fermion-spin fluctuation vertex and
the spin fluctuation propagator.
We find that the fermion self energy and fermion-spin fluctuation vertex are
logarithmically divergent, while the polarization bubble itself which controls
spin fluctuations is not divergent.
We argue that these logarithms sum to a power law and we calculate this power
law to order $1/N$.

\begin{figure}
\centerline{\epsfxsize=4cm \epsfbox{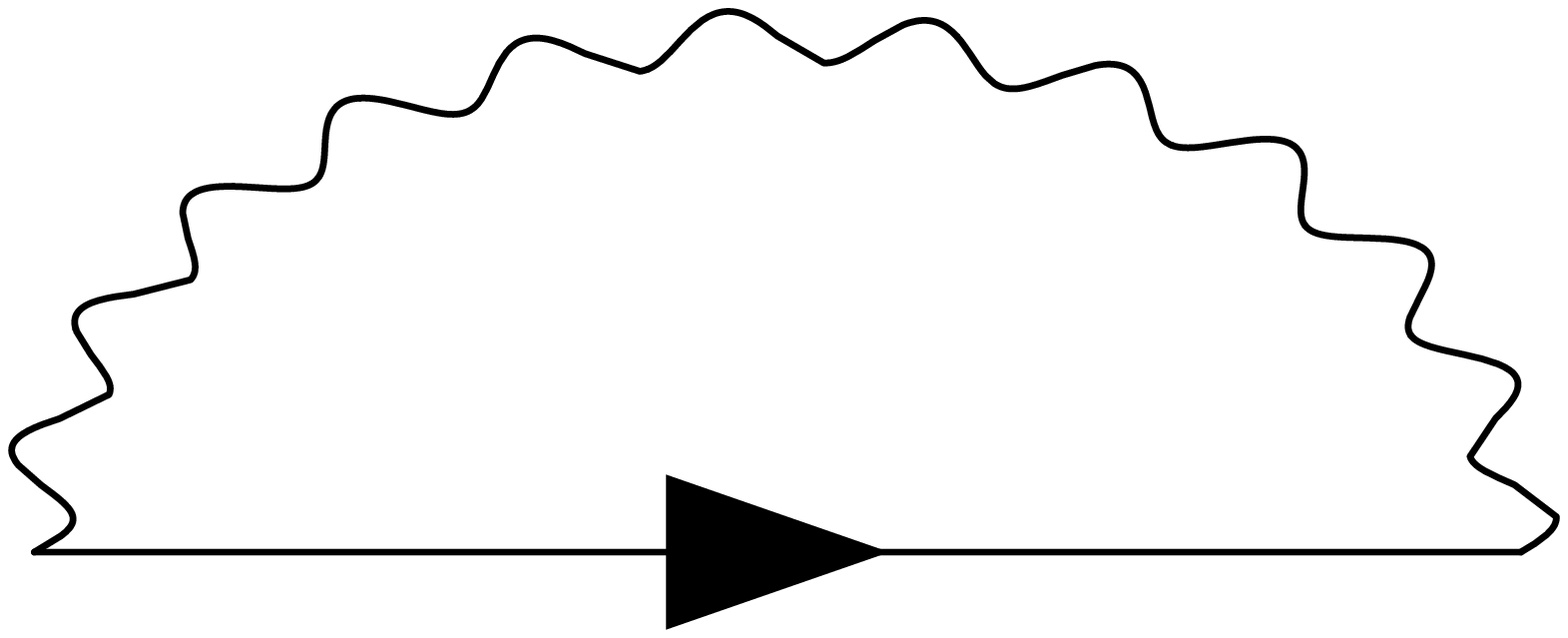}}
\nopagebreak
{
FIG. 4.
Leading contribution to the self energy. Solid line denotes fermion, wavy line
denotes spin fluctuations.
}
\end{figure}
\vspace{0.2in}

We begin with the self energy.
The leading contribution is shown in Fig. 4 and corresponds to
\begin{equation}
\Sigma(\epsilon,{\bf p})= g^2 \int G(\epsilon+\omega,{\bf p'}) D(\omega,{\bf
p-p'}) d\omega d^2 p'
\label{Sigma}
\end{equation}

The form of the fermion Green function implies that the energy of the
scattered electron is small ($\sim \epsilon$), so the frequency transferred to
the spin fluctuations is small and the scattered electron remains near the
Fermi line.
Although $\bf p$ and $\bf p'$ must both be near the Fermi line, the angle
between them may be  large; thus, it will be convenient to parametrize the
momenta $\bf p$ and $\bf p'$ by their polar coordinates $p$, $\theta$, $p'$,
$\theta'$.
Moreover, since the momentum ${\bf p}-{\bf p'}$ transferred to the spin
fluctuation is large, we may neglect the frequency dependence of the spin
propagator except as a lower cut off and estimate the large momentum
${\bf p}-{\bf p'}$ neglecting the small differences $|p|-1$ and $|p'|-1$.
For electrons near the Fermi line the difference
${\bf p}-{\bf p'}$ is always smaller than $2$, so the main contribution to the
spin fluctuation propagator (\ref{D_0}) comes from the term
$\Pi^{(sing)}(\omega,
{\bf G-q})$ in (\ref{Pi_0}), yielding
\begin{equation}
D(\omega, {\bf p}-{\bf p'}) = \frac{4 \pi z}{N \sqrt{
	3\theta^2 + 3\theta'^2 + 2\theta\theta'}}
\label{D_appr}
\end{equation}
The $1/|\theta'|$ dependence of $D(\omega, {\bf p}-{\bf p'})$ at large
$\theta'$ leads to a logarithmic contribution to the self energy and justifies
our assumption that angular deviations are typically large.
The logarithm is cut off by $\Lambda$ which is the largest of $|\epsilon|$,
$p-1$, $\theta^2$.
Substituting Eq. (\ref{D_appr}) into (\ref{Sigma}) we get
\begin{equation}
\Sigma(\epsilon,\theta) = -i z_0 \epsilon \frac{n}{\sqrt{3} N \pi} \ln
(1/\Lambda)
\label{Sigma_l}
\end{equation}
Here $n=3$ is the number of spin components and $z_0$ is the usual Fermi
liquid wave-function renormalization.
We emphasize again that this expression is only correct if the infrared cutoff
$\Lambda$ is larger than $\Delta G$.

Note that $\Sigma$ is a function only of energy and $\theta$, so the
Fermi velocity is not renormalized and the structure (\ref{G}) that we assumed
for the Green  function is not changed.
Note also that the relative value of the renormalization
$\Sigma(\epsilon)/(z\epsilon)$ does not depend on the value of the coupling
constant $g$ or indeed on any other parameter of the theory except for the
number of electron components $N$ and spin components $n$.
Thus, even though $g$ is renormalized by the interaction, this renormalization
is not important for the calculation of $\Sigma(\epsilon)$ and we may
expect that the logarithms sum up to a power law and that the
exponent depends only on $n$ and $N$.
We express the renormalization of the self energy as a scale dependent
wave-function renormalization $z(\Lambda)$ and find
\begin{equation}
z \sim (1/\Lambda)^\alpha
\label{z}
\end{equation}
with $\alpha=n/(N\sqrt{3}\pi)$ in the large $N$ limit.
Note that even at $N=2$, $n=3$ $\alpha \approx 0.27$ is a small number,
suggesting that the leading logarithm approximation is reasonably accurate
even in this case.

\begin{figure}
\centerline{\epsfxsize=4cm \epsfbox{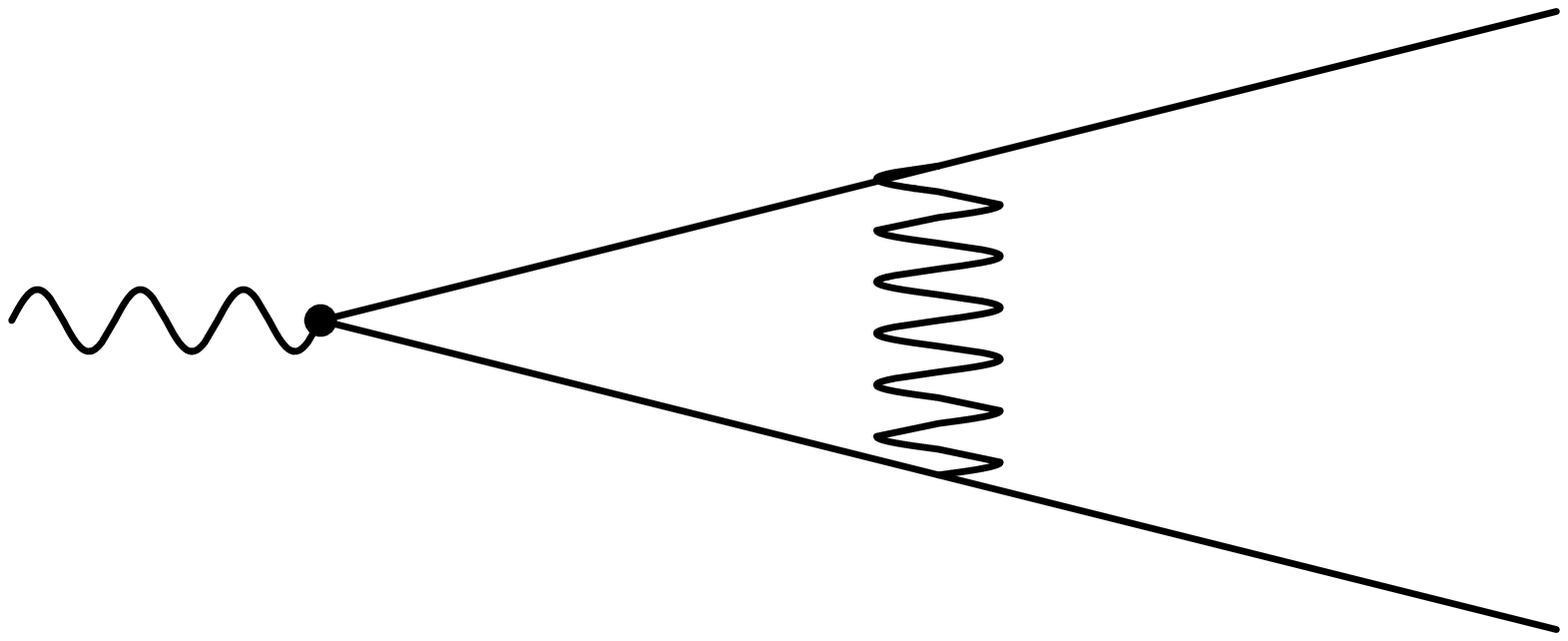}}
{
FIG. 5.
Leading contribution to the vertex renormalization. Solid lines denote
fermions, wavy lines denote spin fluctuations.
}
\end{figure}
\vspace{0.2in}

We now consider the renormalization of the interaction vertex $g$.
At leading order in $1/N$ this is given by the diagram shown in Fig. 5.
The evaluation proceeds differently from the evaluation of the self energy
because the  momenta of the particle and hole are on the opposite sides of the
Fermi line and frequency of the spin fluctuation $\omega$ is not small.
It is convenient to use the cartesian coordinates $p_\parallel$ and $p_\perp$,
introduced in (\ref{G}) and Fig. 1.
The expression corresponding to Fig. 5 is
\begin{eqnarray*}
\frac{\delta g}{g} &=& \frac{2\pi(2-n)}{N} \int
	\frac{d\epsilon dp_\parallel dp_\perp}{(2\pi)^3}
	\frac{1}{i\epsilon - p_\parallel - \frac{p_\perp^2}{2}}
	\frac{1}{i\epsilon + p_\parallel - \frac{p_\perp^2}{2}}
\\
	&\times&
	\frac{1}{Re\left[
		\sqrt{i\omega+\frac{p_\perp^2}{4}+p_\parallel}+
		\sqrt{i\omega+\frac{p_\perp^2}{4}+p_\parallel} \right]}
\end{eqnarray*}
This integral is logarithmic.
The coefficient of the logarithm may be obtained by scaling $\epsilon$ and
$p_\parallel$ by $p_\perp^2$ and evaluating the integral over the rescaled
$\epsilon$ and $p_\parallel$ numerically.
We find
\begin{equation}
\frac{\delta g}{g} = a \frac{n-2}{N \pi} \ln(1/\Lambda)
\label{delta_g}
\end{equation}
with $a \approx 0.75$.
As in the case of the self energy this expression exponentiates, leading to
\begin{equation}
g(\Lambda) = (1/\Lambda)^{\beta}g_0
\label{g}
\end{equation}
with $\beta=a \frac{n-2}{N \pi}$, which in the physically relevant case $n=3$,
$N=2$ becomes $\beta \approx 0.08$ so the corrections to the vertex are very
small and we may assume that the one loop approximation of the vertex
corrections  is reasonably accurate in the physical situation.

Note that both exponents $\alpha$ and $\beta$ depend only on $n$ and $N$ but
not on any other parameter.
This can be also seen directly from the action (\ref{A}) because one can
always scale away the interaction constant $g$ and  the wave function
renormalization $z$ changing the scales of the $S$-fields and frequencies.
This shows that the effective charge controlling the RG flow depends only
on the parameters $n$ and $N$ and is not renormalized.

We finally consider the renormalization of the spin fluctuation propagator
$D(\omega, q)$.
To order $1/N$ there are two diagrams; a self energy and a vertex
correction.
Power counting shows that the frequency and momentum dependent terms in
$\Pi(\omega, q)$ acquire no additional renormalization beyond the one imposed
by the momentum dependent $z$ and $g$, so $D(\omega, q)$ is given by Eq.
(\ref{D}) with  $z$ and $g$ replaced by their running values.
However, there are logarithmic contributions to the mass $\Delta_0$.
These come from the diagrams shown in Fig. 6.

\begin{figure}
\centerline{\epsfxsize=6cm \epsfbox{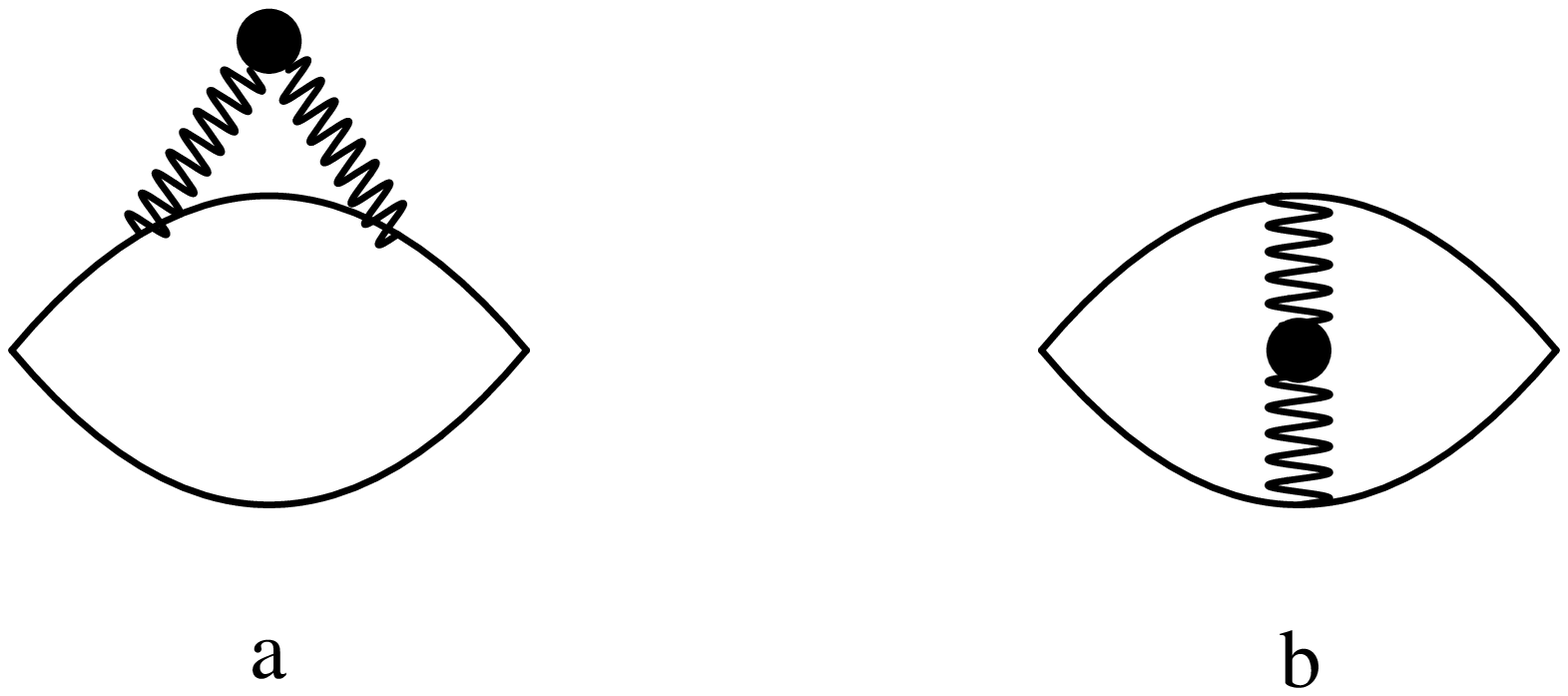}}
{
FIG. 6.
Leading contribution to the mass, $\Delta$, renormalization. Solid line denotes
fermion, wavy lines denote spin fluctuations.
Heavy dot denotes mass operator $\Delta$.
}
\end{figure}
\vspace{0.2in}

The analytic expression corresponding to the diagram in Fig 6a is
\begin{eqnarray}
\frac{\delta \Delta}{\Delta} &=& \int G(\epsilon,{\bf p}+{\bf Q})
	G(\epsilon,{\bf p})^2
\nonumber \\
&\times& G(\epsilon-\omega,{\bf p}-{\bf q}) D^2(\omega,{\bf q})
\frac{d^2p d^2q d\omega d\epsilon}{(2\pi)^6}
\label{delta_Delta1}
\end{eqnarray}
The self energy part of this diagram (second line in (\ref{delta_Delta1}))
has an ultraviolet divergence which leads to a trivial shift of the fermion
chemical potential which we subtract and a logarithmic divergence which we
obtain by performing the integrations in the following order.
We integrate first over $p_\parallel$, then over $\epsilon$ and finally over
$p_\perp$, obtaining an integral over $\omega$, $k_\perp$ and $k_\parallel$.
We then find that $k_\perp$ can be scaled out of the integrals, leading to
\begin{equation}
\frac{\delta \Delta}{\Delta} = - \frac{n}{N} c_a \int
	\frac{d k_\perp}{|k_\perp|} = - \frac{n}{N} c_a \ln(1/\Lambda)
\end{equation}
Here the coefficient
\begin{eqnarray}
c_a &=& \pi^2 \int \frac{dx d\tilde{\omega}}{\left[ Re\left(
	\sqrt{\frac{1}{4}+x+i\tilde{\omega}} +
	\sqrt{\frac{1}{4}-x+i\tilde{\omega}}
	\right) \right]^2}
\nonumber \\
  &\times& Im\left( \frac{\zeta^{-1} \mbox{sgn}(\tilde{\omega})}
		{2i\tilde{\omega}-2x+
	\zeta \mbox{sgn} \tilde{\omega}}
	\right)
\nonumber \\
  &\approx& 0.20
\end{eqnarray}
where $\zeta = \sqrt{4i\tilde{\omega}-4x-1}$.

The contribution of the diagram shown in Fig. 6b may be evaluated similarly,
but in this case no subtraction is necessary.
We find ultimately a logarithmic divergence from the last $q_\perp$ integral.
As was the case for the diagram shown in Fig. 6a, the coefficient of the
logarithm is independent of $g$ and $z$:
\begin{equation}
\frac{\delta \Delta}{\Delta} = \frac{n-2}{N} c_b \ln(1/\Lambda)
\end{equation}
with $c_b \approx 0.45$.
By combining the results of these two diagrams and integrating the resulting
scaling equation we find
\begin{equation}
\Delta(\Lambda) = \Delta_0 \Lambda^{\eta}
\label{Delta(Lambda)}
\end{equation}
with exponent $\eta=2\frac{c_a n - c_b (n-2)}{N} + O(1/N^2)$.
Evaluating this formula at $N=2$ gives $\eta \approx 0.15$.
The formula for $\Delta(\Lambda)$ implies that the mass term becomes important
at a scale set by equating the renormalized kinetic term to the renormalized
mass term, i.e. at
\begin{equation}
g^2(\Lambda) \frac{1}{z(\Lambda)} \sqrt{z(\Lambda)\Lambda} = \Delta_0
\Lambda^{\eta}
\end{equation}
Solving for the scale $\Lambda$ we get
\begin{equation}
\Lambda_\Delta= \Delta_0^{\frac{2}{1+\alpha-4\beta}}
\label{Lambda}
\end{equation}
This result is meaningful only if $\Lambda_\Delta > \Delta G$.
At energy smaller than $\Lambda$ the infrared divergences are absent and
the renormalization flow stops.
One expects that $\Delta_0$ is linear in some external control parameter such
as pressure (which would vary the interaction constant);
thus if $p-p_c > \Delta G$, we would expect that $\chi(0,{\bf Q})$ would vary
with pressure as
\begin{equation}
\chi(0,{\bf Q}) \propto (p-p_c)^{-\frac{2\eta}{1+\alpha-4\beta}}
\label{chi_pr}
\end{equation}

\section{Scaling at small momenta}

For small energy and momentum the splitting, $\Delta G$,  between the peaks
in $\Pi(0,{\bf q})$ becomes important.
To treat this regime it is convenient to expand $\Pi(0,{\bf q})$ for momenta
small compared to $\Delta G$.
It is also convenient to measure parallel momenta in the units of $\Delta G$
and frequencies in the units of $z(\Delta G) \Delta G$.
In these units the fluctuation propagator,
\begin{equation}
D(\omega,{\bf q}) = \frac{2\pi}{ N\left[ Re
\sqrt{k_\parallel + \frac{k_\perp^2}{4} + i \omega} -
	b\left(k_\parallel - \frac{k_\perp^2}{4}\right) + \Delta \right]},
\label{D1}
\end{equation}
contains a new parameter, $b$, which, as we show below, controls
the renormalization group flow and which itself is renormalized.
In our units the initial value of $b$ is $1/2$.

We analyze the model in the same way as in the previous section.
The spin fluctuation propagator is more singular in the region of large and
negative $k_\parallel + \frac{k_\perp^2}{4}$ when the real part of square root
in Eq. (\ref{D1}) is small.
The more singular propagator leads to infrared divergences which are  stronger
than logarithmic, and, we shall show, to a first order transition.

As before, we may consider the renormalization of the electron self-energy,
the interaction constant and the polarization bubble.
Also as before, the renormalization of the self energy does not affect
the renormalization of other quantities; we do not discuss it further.
Unlike the situation at large momenta, there is no renormalization of the
interaction constant in the leading order in $1/N$ because in the diagram of
Fig. 5 it is not possible to put all fermion lines on the Fermi line and
simultaneously have the wavy line carry momentum close to $\bf Q$.
The leading corrections to both $\Delta$ and $b$ thus come only from
the self energy insertion in  the polarization bubble, as shown in Fig. 6a.

The dominant scattering processes contributing to the electron self-energy are
those in which the electron momentum remains close to the Fermi line.
Note that for such processes the momentum transfer is such that
$k_\parallel + \frac{k_\perp^2}{2} \approx 0$, so the spin propagator is large.
To discuss these processes it is convenient to use the radial and angle
coordinates used in the discussion of the electron self-energy in the previous
section.
It further develops that the dominant contribution comes from  processes in
which the angle $\theta$ pertaining to initial momentum $\bf p$ is small
relative to the angle $\theta'$ pertaining to $\bf p'$.
In this limit we may approximate
\begin{equation}
D(\omega, {\bf q})=\frac{2\pi}{N} \frac{1}{\frac{|\omega|}{|\theta'|} +
	\frac{3}{4} b \theta'^2}
\label{D2}
\end{equation}
Using this expression for $D(\omega, {\bf q})$ and evaluating the diagram
shown in Fig 6a gives
\begin{equation}
\frac{\delta \Delta}{\Delta} = - \frac{2n}{27\pi N} \frac{1}{b}
\frac{\ln(1/\Lambda)}{\Lambda^{1/2}}
\label{delta_Delta2}
\end{equation}

\begin{figure}
\centerline{\epsfxsize=4cm \epsfbox{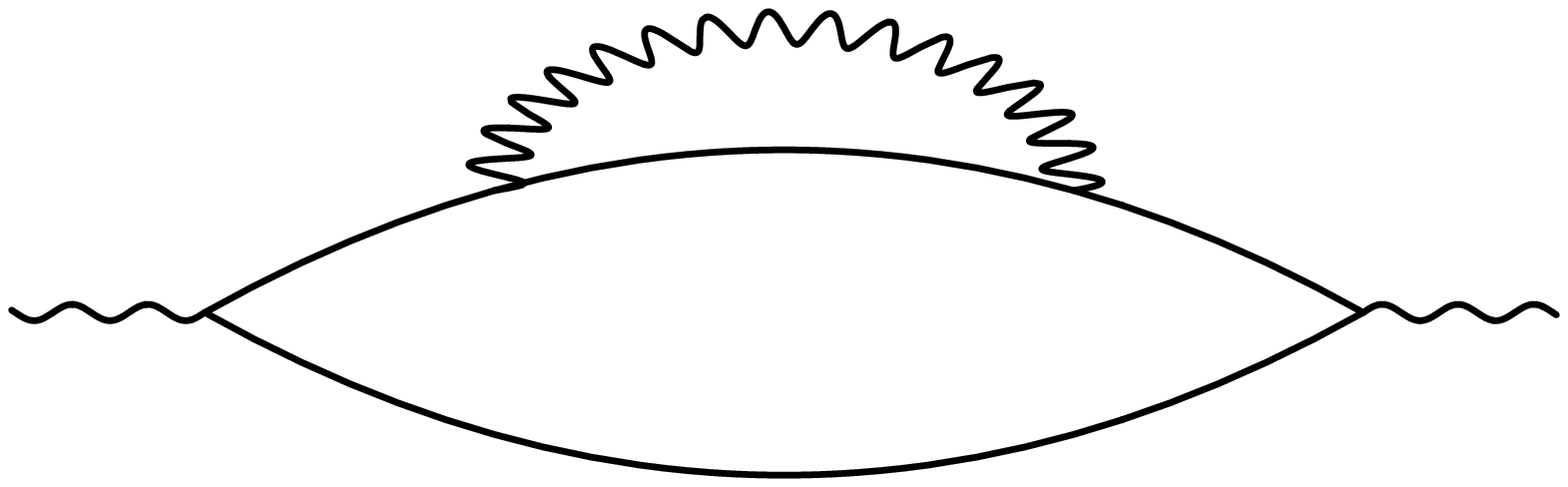}}
{
FIG. 7.
Leading contribution to the renormalization of the spin fluctuation propagator.
Solid lines denote fermions, wavy lines denote spin fluctuations.
}
\end{figure}
\vspace{0.2in}

The leading correction to $b$ comes from the diagram shown in Fig. 7 and may be
evaluated similarly.
We obtain
\begin{equation}
\frac{\delta b}{b} = - \frac{3^{5/6}n}{10 N (2b)^{2/3}}
\frac{1}{\Lambda^{5/6}}
\label{delta_b}
\end{equation}
Equation (\ref{delta_Delta2}) implies that $\Delta$ decreases exponentially as
$\Lambda$ is decreased.
This means that it is not possible to find a self consistent solution along
the lines of Eq. (\ref{Lambda}) for $\Lambda$ less than a number of order $1$.

Arguments originally developed by Brazovskii \cite{Brazovskii} in a slightly
different context show that the minimal value of $\Lambda$ implies a first
order transition.
The physical reason is that fluctuations lead to such a large increase in the
energy of the critical state that at some point it is favorable to
discontinuously open a gap, gaining condensation energy and suppressing
fluctuations.
In the present problem, the fluctuations are so strong that for physical
values of $n$, $N$ and $b$, the first order transition happens almost
immediately as the scale $\Delta G$ is reached and we expect that the
discontinuities in physical quantities are on the scale set by $\Delta G$.

\section{Physical Consequences}

In this Section we describe the implications of our results for observables.
We focus on the intermediate scaling regime discussed in Section III.
The observables are determined by the spin susceptibility which is restoring
units:
\begin{eqnarray}
\chi(\omega,{\bf q}) &=& \frac{g_0^{-2}}{
	g^2(\Lambda) \left[ \Pi(\omega,{\bf q}) +  \Pi(\omega,{\bf Q}-{\bf q})
	\right] + \Delta}
\label{chi_fin} \\
\Pi(\omega,{\bf q}) &=& \frac{N \sqrt{p_0 p_F}}{2 \pi z(\Lambda) v_F}
	{\mbox Re} \sqrt{ \frac{|q|-2p_F}{p_F} +
	\frac{i z(\Lambda) \omega}{v_F p_F}}
\label{Pi_fin}
\end{eqnarray}
Here $g^2(\Lambda)$ and $z(\Lambda)$ are slow power law functions of momentum
and energy; explicit formulas are given in Eqs. (\ref{z},\ref{g}).
We emphasize that these formulae only apply at scales larger than the peak
splitting $\Delta G$.
This form for $\chi(\omega,{\bf q})$ is essentially the RPA form with small
modifications due to the momentum and frequency dependence of $g^2$ and $z$.
The Eq. (\ref{chi_fin}) is written in Matsubara frequencies.
The imaginary part of the analytic continuation of $\chi(\omega,{\bf q})$ is
measurable in neutron scattering experiments.
The actual form of this imaginary part is somewhat complicated, due to the
structure associated with the boundaries of the particle-hole continuum, so we
do not write it here.
We do discuss in more detail the predictions for NMR relaxation rates $1/T_1$
and $1/T_2$, which involve the low frequency limits of the real and imaginary
parts of $\chi(\omega,{\bf q})$ respectively.

The static limit of the real part of $\Pi(\omega,{\bf q})$ is
easily obtained from Eq. (\ref{Pi_fin}) and is
\begin{equation}
{\mbox Re} \Pi(0,{\bf q}) = \frac{N \sqrt{p_0 p_F}}{2 \pi z(T) v_F}
	{\mbox Re} \sqrt{ \frac{|q|-2p_F}{p_F} }
\label{Re_Pi}
\end{equation}

There are different regimes for the imaginary part of $\Pi(\omega,{\bf q})$
at small frequency $\omega \ll T$.
We find more convenient to evaluate $Im \Pi$ directly from the
diagram shown in Fig.
2 using renormalized Green functions and vertices than to analytically
continue equation (\ref{Pi_fin}).
The general expression,
\begin{eqnarray*}
&&\lim_{\omega \rightarrow 0} \frac{{\mbox Im} \Pi(\omega, {\bf q})}{\omega}
	=
\\
&&\int
	{\mbox Im} G_R(\epsilon, {\bf p}+{\bf q})
	{\mbox Im} G_R(\epsilon, {\bf p})
\frac{d^2p d\epsilon g^2(\epsilon,{\bf p}) }
	{(2\pi)^3 2T {\mbox cosh}^2[\epsilon/(2T)]}
\end{eqnarray*}
is obviously dominated by  frequencies of the order of $T$ and momenta
$T/v_F$ so for this calculation we can use $g^2(\Lambda)$ and $z(\Lambda)$
evaluated at $\Lambda=T$.
If $v_F(|q|-2p_F) < z(T)T$ we get
\begin{equation}
\lim_{\omega \rightarrow 0} \frac{{\mbox Im} \Pi(\omega, {\bf Q})}{\omega} =
\frac{N c_1 p_0^{1/2} g^2(T)}{2\pi z^{1/2}(T) v_F^{3/2} T^{1/2}}
\label{Im_Pi_0}
\end{equation}
where $c_1 \approx 0.23$.
If $v_F(|q|-2p_F) > z(T)T$ the integral is cut off by the external momentum and
is
\begin{equation}
\lim_{\omega \rightarrow 0} \frac{{\mbox Im} \Pi(\omega, {\bf q})}{\omega} =
\frac{N p_0^{1/2} g^2(T) }{4\pi v_F^{2} (|q|-2p_F)^{1/2}}
\label{Im_Pi_1}
\end{equation}
We may now calculate the relaxation rates by combining the results above
with the general relation of the real and imaginary part of the
susceptibility to the polarization operator
\begin{eqnarray*}
\chi'(0,{\bf q}) &=& \frac{1}{
	g_0^2 \left[ g^2 {\mbox Re} \Pi(0,{\bf q}) + \Delta \right] }
\\
\lim_{\omega \rightarrow 0} \frac{\chi''(\omega,{\bf q})} {\omega} &=&
\lim_{\omega \rightarrow 0}
	\frac{ g^2 {\mbox Im} \Pi(\omega,{\bf q})}
	{\omega g_0^2 \left[g^2 {\mbox Re} \Pi(0,{\bf q}) +
		\Delta \right]^2  }
\end{eqnarray*}
and inserting the results into the general expressions for relaxation rates
\begin{eqnarray*}
\frac{1}{T_1 T} &=& \sum_q A_q
	\lim_{\omega \rightarrow 0} \frac{\chi''(\omega,{\bf q})}{\omega}
\\
\frac{1}{T_2} &=& \sqrt{ \sum_q \left( A_q \chi'(0,{\bf q}) \right)^2}
\end{eqnarray*}
where $A_q$ is determined by hyperfine couplings.
We
see that the $T_2$ rate behaves as
\begin{equation}
\frac{1}{T_2} = A - B T^{\frac{1}{4} + \frac{5}{4} \alpha - 2 \beta}
\label{T_2}
\end{equation}
or, using our previous results $\alpha \approx 0.27$, $\beta \approx 0.08$
\[
\frac{1}{T_2} = A - B T^{0.05}
\]
Of course, our estimates for $\alpha$ and $\beta$ come from a $1/N$ expansion,
so for the $N=2$ case we may conclude that $1/T_2$ rate is either weakly
divergent or non-divergent but with anomalously rapid  $T$-dependence at low
temperatures.
Evaluating $1/T_1T$ similarly we find
\begin{equation}
\frac{1}{T_1 T} = C T^{-\frac{3}{2} \alpha + 2 \beta} \approx C T^{-0.25}
\label{T_1}
\end{equation}
i.e. we expect the $1/T_1T$ rate to be weakly divergent assuming (as was found
at large $N$) $\alpha > 4 \beta/3$.

\begin{figure}
\centerline{\epsfxsize=6cm \epsfbox{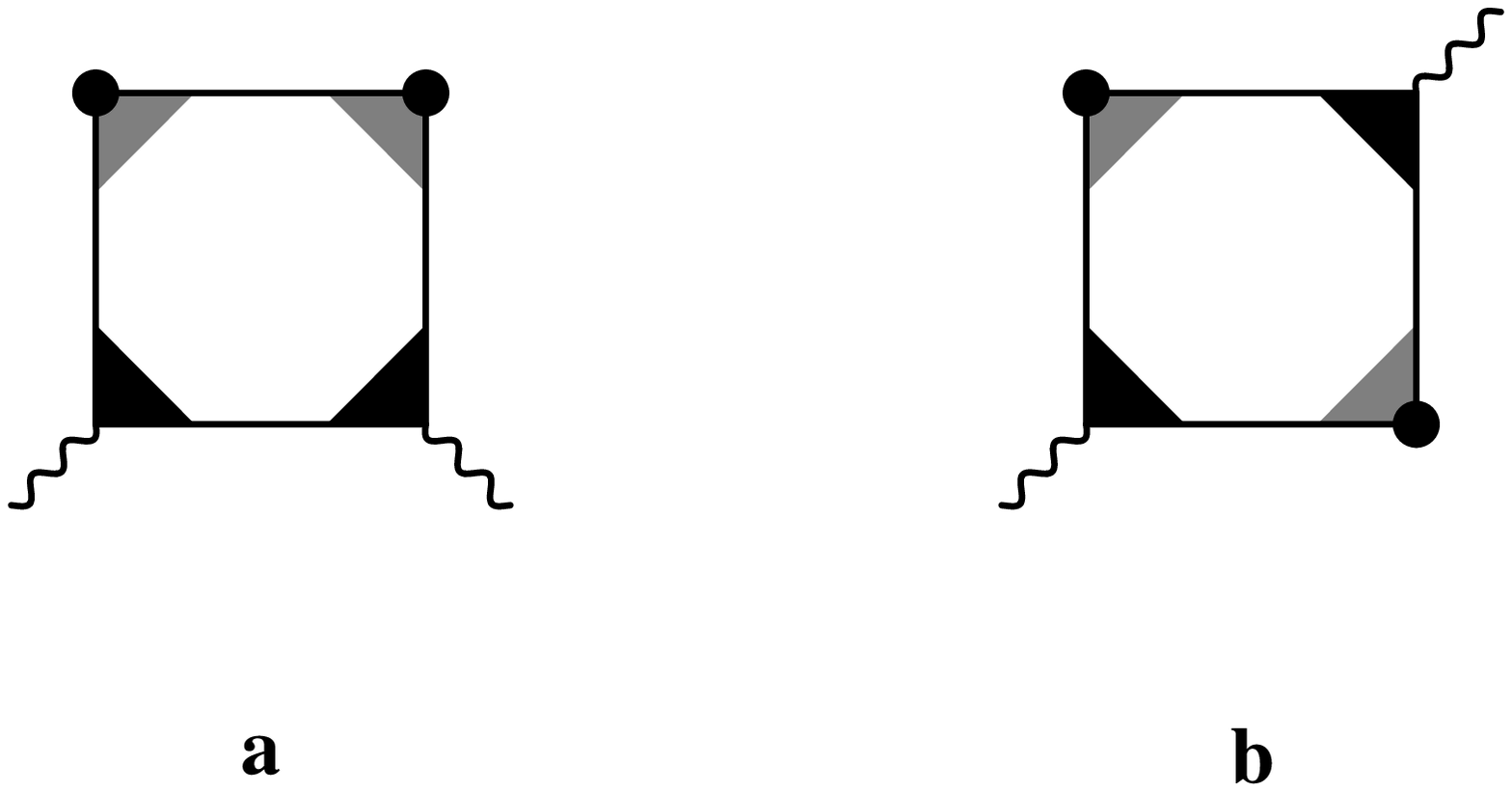}}
{
FIG. 8.
Leading contributions to the renormalization of the uniform susceptibility.
Solid lines denote fermions, wavy lines denote spin fluctuations, solid dot
denotes coupling $g_e$ to the external magnetic field.
}
\end{figure}
\vspace{0.2in}

Proximity to the antiferromagnetic transition has also an effect on the
uniform susceptibility.
We have shown elsewhere \cite{chipaper} that the leading low-$T$ behavior of
$\chi({\bf q},0)$ in the limit
${\bf q} \rightarrow 0$
is given by
\begin{equation}
\chi_U = \lim_{{\bf k} \rightarrow 0} \chi(0,{\bf k}) =
\sum_{q,\nu} B (k,\nu) D(\nu,{\bf q})
\label{chi(0,0)_eq}
\end{equation}
where the coefficient $B$ is given by the diagrams in Fig. 8, in which fermion
mediates interaction between spin fluctuations and external magnetic field.
We denote the bare vertex coupling fermions to the external magnetic field by
$g_e$.

In order to calculate $B$ we use renormalized fermion Green functions and
renormalized fermion-spin-fluctuation vertices $g$.
The coupling to the uniform external field is not renormalized; this follows
mathematically from the fact that in any vertex correction diagram the poles
in the Green function lines are in the same half-plane, so in the limit of
vanishing external frequency the low-energy contribution vanishes.
Alternatively, one can show that the absence of logarithmic $g_e$
renormalization follows
from the fact  that the logarithmic renormalization given in (\ref{z}) does not
affect the fermion density of states.
Estimating the diagrams in Fig. 8 we get
\begin{equation}
B \sim z^{-5/2}(\epsilon) g^2(\epsilon) \epsilon^{-3/2}
\end{equation}
Combining this with our result for $D(\omega,{\bf q})$ gives
\begin{equation}
\chi_U \sim c_1 + c_2 T^{\frac{1+\alpha}{2}}
\label{chi_U}
\end{equation}
Finally, we note that this scaling regime only exists for energy, temperature
and momenta cutoff greater than $\Delta G$.
As soon as the relevant scales fall below $\Delta G$, the $T=0$ transition
becomes first order.
This transition happens at some value, $\Delta_0^c$ of the controlling
parameter $\Delta$.
At small but non-zero temperatures the transition remains a first order but it
occurs at a somewhat different value of $\Delta(T)$.
We may estimate this transition line from the argument that if $\Delta <
\Delta_0^c$, then the ordered phase has lower energy than the disordered
phase, so $\Delta E = E_0 (\Delta_0^c - \Delta_0)$.
However, the disordered phase has greater entropy, because there is no gap on
the Fermi line; thus $\Delta F = - T^2 S_0$.
Equating the two gives
\begin{equation}
T_{c1} = T_0 (\Delta_0^c - \Delta)^{1/2}
\label{T_c1}
\end{equation}
We expect that at scales greater than $\Delta G$ the line of first order
transitions terminates at a critical point, $T_c$.
We estimate $T_c (T_c/E_F)^{\alpha} \sim v_F \Delta G$.

\section{Conclusion}

We have presented arguments suggesting that the $2p_F$ density wave transition
is first order in $d=2$ spatial dimensions.
By way of conclusion we place our results in the context of quantum critical
phenomena and of theories of strongly correlated electrons, and discuss some
implications of our conclusions for the physics of high $T_c$ superconductors.

One may divide quantum critical phenomena in metals into two classes: those in
which fluctuations of the order parameter are strongly coupled to the
particle-hole continuum (and in particular may decay into a particle-hole
pair) and those in which the coupling to fermions is irrelevant.
If the coupling to fermions is irrelevant, then the critical phenomena are
described by a theory of propagating bosons which may be studied by
conventional methods \cite{Vaks,Weichman,Sachdev}.
If decay into a particle-hole pair is possible, then in most cases it is still
possible to describe the critical phenomena by a theory of bosons, albeit with
overdamped dynamics \cite{Hertz,Millis,AILM}.
The description in terms of a purely bosonic theory is possible in these cases
because the effect of the critical fluctuations on the fermions is small
enough so renormalization of the fermions does not feed back into the
properties of critical fluctuations.

The one exceptional case is the two dimensional $2p_F$ transition.
Here the large phase volume available for scattering across the Fermi line
makes the critical fluctuations have a strong effect on the electrons;
further the   $2p_F$ singularities in electron response functions mean that
electrons near the Fermi line which are strongly affected by critical
scattering have a large effect on the critical fluctuations.
This physics leads to apparently intractable difficulties in the bosonic model
generated by formally integrating out the electrons.
Specifically, in the resulting bosonic model nonlinear terms of all orders
have divergent coefficients and are all relevant in renormalization group
sense.
Therefore, we used a model in which both electrons and spin fluctuations are
retained.
We assumed that the bare propagators and the susceptibilities have the Fermi
liquid form and that the Fermi line is not straight near the points connected
by the ordering wave vector $\bf Q$.
The structure of the $2p_F$ singularities enabled us to construct a
renormalization group transformation under which we could treat fermions and
spin fluctuations on the same footing.

The solution of the resulting renormalization group equations implies that the
transition is generically first order, because the critical fluctuations are
so strong that they completely suppress the second order transition.
There is one special case where the transition is not first order.
If twice the ordering vector $\bf Q$ is commensurate with a reciprocal lattice
vector, $\bf G$, i.e. $2{\bf Q}={\bf G}$, then the spin fluctuation propagator
is less singular, the fluctuations are weaker and the transition turns out to
be second order and characterized by the exponents which we calculate in a
$1/N$ expansion.
If $|2{\bf Q} - {\bf G}|$ is small, the $T=0$ transition is ultimately first
order but a broad scaling regime exists.

Our explicit calculations were performed for a model of a spin density wave
transition.
Most of our results carry over to the charge density wave case, if the number
of spin components, $n$, is set to 1.
There is one important caveat.
If the system is sufficiently symmetric, e.g. if the Fermi line is circular,
cubic terms (forbidden in the SDW case by time reversal) may exist in Landau
free energy.
Cubic terms lead to the first order transition but their presence would also
complicate the analysis given above.

We now place our results in the context of theories of interacting electrons
in two spatial dimensions.
To produce a quantum phase transition one must increase an interaction
parameter to a critical value which is generally large.
One must therefore consider the effect of the interaction on the non-critical
state of the electrons.
There are two possibilities: one is Fermi liquid theory in which it is assumed
that perturbation theory may be resummed to all orders.
Within this assumption one may show that away from any critical point the low
energy properties are not qualitatively changed from those of free electrons,
so the transition takes place against Fermi liquid background and
one may use the theory given here to calculate the extra effects due to
criticality.

An alternative possibility is that Fermi liquid theory is not a correct
description of the low energy properties even far from criticality.
For example, it has been argued that in strongly correlated two dimensional
models the spin degrees of freedom form a ``spin liquid''
\cite{spin-liquid,Lee}.
A spin liquid possesses a Fermi surface, spin $1/2$ fermionic excitations with
constant density of states at low energies and a particle-hole continuum but
the fermions interact via a singular interaction which, among other things,
changes substantially $2p_F$ singularities \cite{AIM}.
The singular interaction causes anomalous temperature dependence
of susceptibility and NMR relaxation rates even away from the
criticality \cite{AILM} and changes the critical properties.
The spin-liquid to antiferromagnet transition has been studied elsewhere
\cite{AILM}.

The high temperature superconductors are quasi two dimensional materials with
spin dynamics which have been claimed to be controlled by a quantum critical
fixed point.
It is still controversial whether Fermi liquid theory is the correct starting
point for a description of the low energy electron physics.
It is therefore interesting to compare our theoretical results to the known
magnetic properties of high $T_c$ materials.

For a two dimensional Fermi liquid the bare polarizability is so strongly
peaked at $2p_F$ that it is most natural to assume that the transition occurs
at $Q=2p_F$.
This is consistent with neutron scattering data on $La_{2-x}Sr_xCuO_4$, in
which strong and temperature dependent peaks were observed.
These peaks are centered at $x$-dependent wavevectors ${\bf Q}(x)$ which are
claimed to be $2p_F$ wavevectors of the LDA band structure \cite{Littlewood}.
Therefore if a Fermi liquid picture is appropriate, it is natural to expect
our results to describe experiments on $La_{2-x}Sr_xCuO_4$.

In fact, they do not.
There is no sign of a first order transition; the magnetic properties
apparently evolve smoothly with doping.
It is conceivable that disorder due to random positions of the $Sr$ dopants
masks the first order transition.
However, for all $Sr$ concentrations including $x=0.14$ there is a wide
temperature regime in which the NMR relaxation rates and uniform susceptibility
vary with temperature.
Roughly, the copper $1/(T_1T)$ and $1/T_2$ are inversely proportional to $1/T$
\cite{T_1,T_2} while $\chi_U \sim A + B T$ \cite{chi_U}.
These temperature dependences are not consistent with our results, Eqs
(\ref{T_1},\ref{T_2},\ref{chi_U}).
We conclude that the magnetic properties of $La_{2-x}Sr_xCuO_4$ are not well
described within a Fermi liquid approach.
Two alternatives have been proposed: one is that the fermions are in
``spin-liquid'' regime described above \cite{AILM}; another is that the
critical behavior
is due to propagating spin waves only weakly coupled to the electrons
\cite{Chubukov,Sokol}.

Another class of materials to which our results might be relevant are the low
dimensional organics such as TTF-TCNQ.
These materials have strongly anisotropic transfer integrals $t_a \gg t_b \gg
t_c$, leading typically to an open Fermi line, but with non-negligible
curvature.
The curvature implies that the physics of these materials is not strictly
one-dimensional and makes it possible that the theory developed here is
relevant and explains the experimental observation that the spin density wave
transition produced by lowering the temperature is weakly first order
\cite{Clark94}.

{\em Note added}.  After this paper was completed we received a preprint
from A. V. Chubukov analysing the spin density wave transition
with ${\bf Q} = {\bf G}/2$ but ${\bf Q} \neq 2p_F$.  He found
a logarithmic renormalization very similar to the one we found
in the intermediate scaling regime discussed in section III and
showed that this implies that the exponents characterizing
the critical point he analysed differ from
the exponents characterizing the general case in
which ${\bf Q} \neq {\bf G}/2$
and ${\bf Q} \neq 2p_F$ analysed by previous authors \cite{Hertz,Millis}.

\end{multicols}


\begin{references}
\bibitem{remark}
The $2p_F$ wavevectors were referred to in previous work \cite{Millis} as
``maximal spanning vectors'' but this terminology is misleading because the
essential feature (parallel tangents) can occur even if $\bf Q$ is not a
maximal spanning vector.
\end{references}
\end{document}